\documentclass[prl,preprint,floatfix,showpacs,superscriptaddress]{revtex4}

\usepackage[latin1]{inputenc}
\usepackage{amsmath}
\usepackage{mathrsfs }

\usepackage[usenames,dvipsnames]{color}
\usepackage{amsfonts}

\usepackage{amssymb,latexsym} 

\usepackage{enumerate}

\usepackage{graphicx}

\usepackage{hyperref}
\usepackage{multirow}
\newcommand{\beq}[1]{\begin{equation}\label{#1}}
\newcommand{\eeq}{\end{equation}}

\begin{document}
\title{
Log-log Convexity of Type-Token Growth in Zipf's Systems
}
\author{Francesc Font-Clos}
\author{\'Alvaro Corral}
\affiliation{Centre de Recerca Matem\`atica,
Edifici C, Campus Bellaterra,
E-08193 Barcelona, Spain.
} 
\affiliation{Departament de Matem\`atiques,
Facultat de Ci\`encies,
Universitat Aut\`onoma de Barcelona,
E-08193 Barcelona, Spain}

\begin{abstract}

It is traditionally assumed that Zipf's law implies 
the power-law growth of the number of different elements
with the total number of elements in a system 
- the so-called Heaps' law.
We show that a careful definition of Zipf's law
leads to the violation of Heaps' law in random systems,
and obtain alternative growth curves.
These curves fulfill universal data collapses
that only depend on the value of the Zipf's exponent. 
We observe that real books behave very much in the same way as random
 systems, despite the presence of burstiness in word occurrence.
We advance an explanation for this unexpected correspondence.
\end{abstract}

\date{\today}

\maketitle

A great number of systems in social science, economy, 
cognitive science, biology, and technology have been proposed to 
follow Zipf's law 
\cite{Newman_05,Clauset,Adamic_Huberman,Furusawa2003,Axtell,Serra_scirep}.
All of them have in common that they are composed by some ``elementary''
units, which we will call tokens, and that these tokens 
can be grouped into larger, concrete or abstract entities, called types.
For instance, if the system is the population of a country, the tokens are its citizens, 
which can be grouped into different concrete types given by the cities where they live
\cite{Malevergne_Sornette_umpu}.
If the system is a text, each appearance of a word is a token, 
associated to the abstract type given by the word itself \cite{Zanette_book}.
Zipf's law deals with how tokens are distributed into types,
and can be formulated in two different ways,
which are generally considered as equivalent
\cite{Newman_05,Adamic_Huberman,Zanette_book,Lu_2010}.

The first one is obtained when counting the number of tokens associated
to each type, which are represented by their rank in the list of counts; 
if a (decreasing) power law holds between counts and ranks,
with an exponent close to one,
this indicates the fulfilment of Zipf's law.
The second version of the law arises when counting the number of types with 
a given value of the number of counts
(this is the distribution of counts) 
and this yields a (decreasing) power law
with exponent around two.
However, in general, the fulfilment of Zipf's
law has not been tested with rigorous statistical methods 
\cite{Clauset,Corral_Boleda};
rather, researchers have become satisfied with just qualitative resemblances
between empirical data and power laws.
In part, this can be justified by the difficulties of obtaining clear statistics
from the rank-count representation, in particular for high ranks (that is, for rare types),
and also by poor methods of estimation of probability distributions \cite{Clauset}.

An important fact in
most Zipf-like systems is that these 
present a temporal order.
And whereas Zipf's law reports a static property of these systems
(as it is not altered under re-ordering of the data),
a closely related statistics can unveil some of the dynamics.
This is the type-token growth curve,
which counts the number of types, $v$, as a function of the number of tokens, $\ell$,
as a system evolves, i.e., as citizens are born or a text is being read.
Note that $\ell$ is a measure of system size (as system grows)
and $v$ is a measure of richness or diversity of types
(with the symbol $v$ borrowed from linguistics, where it stands for 
the size of the vocabulary).

It has long been assumed that Zipf's law implies also a power law
for the type-token growth curve, i.e., 
\begin{equation}
v(\ell) \propto \ell^{\,\alpha},
\end{equation}
with exponent $\alpha$ smaller than one,
and this is referred to as  Heaps' law in general or Herdan's law in quantitative linguistics
\cite{Heaps_1978,Baeza_Yates00,Baayen}.
Indeed, Mandelbrot \cite{Mandelbrot61} and the authors of Ref. \cite{Leijenhorst_2005}
obtain Heaps' law when drawing independently tokens from a Zipf's system.
Baeza-Yates and Navarro \cite{Baeza_Yates00}
argue that, if both Zipf's law and Heaps' law are fulfilled, their exponents are connected.
A similar demonstration, using a different scaling of the variables, 
is found in Ref. \cite{Kornai2002}, and with some
finite-size corrections in Ref. \cite{Lu_2010}.
Other authors have been able to derive Heaps' law from Zipf's law using 
a master equation \cite{Serrano} or
pure scaling arguments \cite{FontClos_Corral}.
Alternatives to Heaps' formula are listed in Ref. \cite{Wimmer_Altmann},
but without a theoretical justification.

However, even simple visual inspection 
of the log-log plot of empirical type-token growth curves shows
that Heaps' law is not even a rough approximation 
of the reality.
On the contrary, a clear convexity (as seen from above)
is apparent in most of the plots
(see, for instance, some of the figures in \cite{Lu_2010,Sano2012,Bernhardsson_2011}).
This has been attributed to the fact that the asymptotic regime
is not reached or to the effects of the exhaustion of the number
of different types \cite{Lu_2013}.
Nevertheless, the effect persists in very large systems, composed by many millions of tokens,
and where the finiteness of the number of available types is questionable
\cite{FontClos_Corral}.

In the few reported cases where there seems to be a true 
power-law relation between 
number of tokens and number of types,
as in Ref. \cite{Kornai2002},
this turns out to come from a related but distinct statistics.
Instead of considering
the type-token growth curve in a single, growing system ($v(\ell)$ for $\ell=1\dots L$), 
one can look for the total type-token relationship in a collection or ensemble of 
$\mathcal{N}$ systems ($V_j$ versus $L_j$, for $j=1\dots\mathcal{N}$,
with $V_j=v(L_j)$),
see also Refs. \cite{Serrano,Petersen_scirep,Gerlach_Altmann,Gerlach}. 
We are, in contrast, interested in the type-token relation of a single growing system.

The fact that Heaps' law is so clearly violated 
for the type-token growth, 
given that this law follows directly from Zipf's law,
casts doubts on the very validity of the latter law.
But one may notice that,
although the two versions of Zipf's law mentioned above are
usually considered as equivalent, they are only asymptotically equivalent
in the limit of very high counts \cite{Mandelbrot61,Heaps_1978,Baayen}. 
However, the type-token growth curve emerges mainly from the statistics of the rarest types,
for every $\ell$, 
as it is only when a type appears for the first time that it contributes
to the growth curve \cite{FontClos_Corral}, and these are precisely the types for
 which the usual description in terms of the rank-count representation becomes
 problematic.
So, the election of which is the form of Zipf's law that one considers to hold true
becomes crucial for the derivation of the type-token growth curve and
the fulfilment of Heaps' law or not.

Although most previous research has focused in Zipf's law in the rank-count representation, 
\emph{i.e.}, the first version mentioned above,
we argue that it is the second version of the law, 
that of the distribution of counts,
the one that becomes relevant to describe the real type-token growth curve, 
at least in the case of written texts.
Indeed, let us notice that the previous derivations of Heaps' law
were all based on the rank-count representation
\cite{Mandelbrot61,Leijenhorst_2005,Baeza_Yates00,Kornai2002,Lu_2010,Serrano,FontClos_Corral};
therefore, the violation of Heaps' law for real systems 
invalidates the (exact) fulfilment of Zipf's law for the rank-count representation.

In contrast, when the viewpoint of Zipf's law for the distribution of counts
is adopted, 
we prove that Heaps' law cannot be sustained for random systems
and we derive an alternative law,
which leads to ``universal-like'' shapes of the rescaled type-token 
growth curves, 
with the only dependence on the value of the Zipf's exponent.
Quite unexpectedly, 
our prediction for random uncorrelated systems holds
very well also for real texts.
We are able to explain this effect despite the significant clustering or burstiness
of word occurrences \cite{Corral_words,Motter},
due to the special role that the first appearance in a text of a type plays,
in contrast to subsequent appearances.

Let us consider a Zipf's system of total size $L$,
and a particular type with overall number of counts $n$;
this means that 
the complete system contains $n$ tokens of that type
(and then $L$ is the sum of 
counts of all types, $L=\sum_i n_i$).
In fact, Zipf's law tells us that there can be many types with the same counts $n$,
and we denote this number as $N_L(n)$.
Quantitatively, in terms of the distribution of counts,
Zipf's law reads
\begin{equation}
N_L(n) \propto \frac 1 {n^\gamma},
\label{Zipfnonorm}
\end{equation}
for $n=1,2,\dots$
with the exponent $\gamma$ close to 2.
Note that $N_L(n)$ is  identical, except for normalisation, to the 
probability mass function of the number of counts.

For a part of the system of size $\ell$, with $\ell \le L$,
the number of types
with $k$ counts will be
$N_\ell(k)$.
The dependence of this quantity with the global $N_L(n)$ will be
computed for a random system,
which is understood as a sequence of tokens 
where these are taken at random from some underlying distribution.
The $N_L(n)$ words with number of counts $n$ in the whole system will lead,
on average, to
$N_L(n) h_{k,n}$ types with counts $k$ in the subset,
with $k\le n$ and $h_{k,n}$ given by the hypergeometric distribution, 
\begin{equation}
h_{k,n}= 
\frac{\binom{n}{k} \binom{L-n}{\ell-k}}{\binom{L}{\ell}}.
\end{equation}
This is the probability to get $k$ instances of a certain type when drawing, 
without replacement,
$\ell$ tokens from a total population of $L$ tokens
of which there are $n$ tokens of the desired type.
The dependence of $h_{k,n}$ on $\ell$ and $L$ is not explicit,
{to simplify the notation}.
The {average} number of types with $k$ counts in the subset of size $\ell$
will result from the sum of $N_L(n) h_{k,n}$ for all $n\ge k$, i.e.,
\begin{equation}
N_\ell(k)=\sum_{n\ge k} N_L(n) h_{k,n}.
\label{primera}
\end{equation}
We will use this relationship between $N_\ell(k)$ and $N_L(n)$ to derive the type-token growth curve.

For a subset of size $\ell$ we will have that, out of the total $V$ types,
$v(\ell)$ will be present whereas $N_\ell(0)$ will not have appeared
(and so, their number of counts will be $k=0$); 
therefore, $v(\ell)=V-N_\ell(0)$,
and substituting Eq. (\ref{primera}) for $k=0$ and using that $N_L(0)=0$, then,
\begin{equation}
v(\ell)=V-\sum_{n\ge 1} N_L(n) h_{0,n}.
\label{segunda}
\end{equation}
This  
formula relates the type-token growth curve 
with the distribution of counts in a random system,
where it is exact, {if we interpret $v(\ell)$ as an average
over the random ensemble.}
We now show that a power-law distribution of type counts
does not lead to a power law in the type-token growth curve, 
in other words, Zipf's law for the distribution of counts does not lead to Heaps' law,
in the case of a random system.
\begin{figure}
\includegraphics[width=1.05\columnwidth]{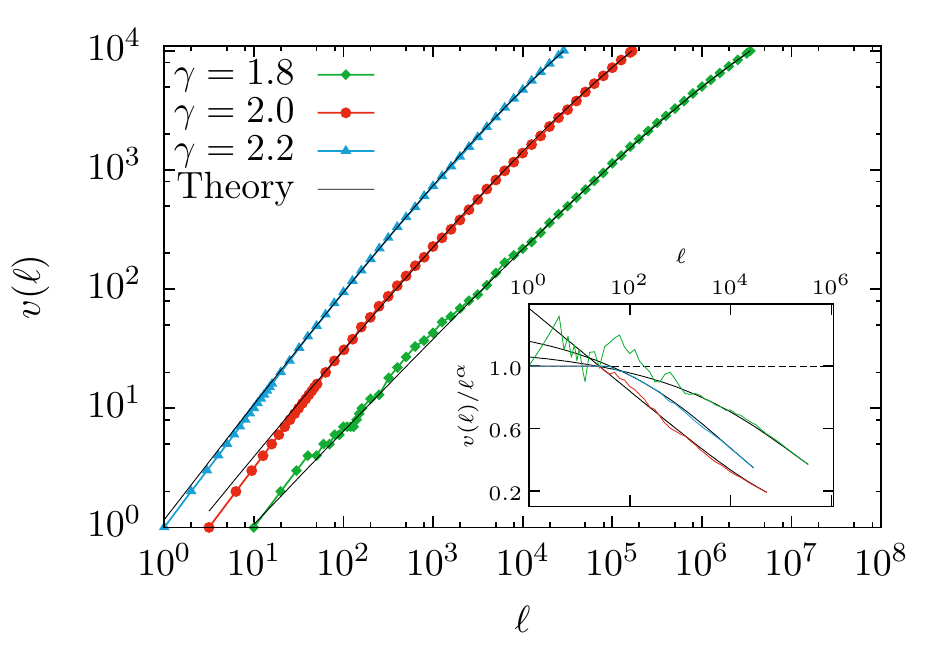}
\caption{\textbf{Main: }Type-token growth curve $v(\ell)$ for three random systems with 
number of counts drawn from a discrete power-law distribution $N_L(n) \propto n^{-\gamma}$, and $\gamma=1.8$ (green diamonds), $2.0$ (red circles) and $2.2$ (blue triangles). The black lines correspond to our theoretical predictions, Eq.
(\ref{theory gamma more 2}) for $\gamma \geq 2$ and Eq.  
(\ref{theory gamma less 2}) for $\gamma < 2$ (plotted with the help of the GSL libraries). 
No average over the reshuffling procedure is performed. 
Curves are consecutively shifted by a factor of $\sqrt 10$ in the $x$-axis. 
\textbf{Inset:} The ratio $v(\ell)/\ell^\alpha$ is displayed, with $\alpha=\min\{1,\gamma-1\}$, showing that an approximation of the form $v(\ell) \propto \ell^\alpha$ is too crude.
}
\label{fig random}
\end{figure}

First, taking advantage of a symmetry of the hypergeometric distribution
and making an approximation for $n\ll L$, 
 the ``zero-success'' probability $h_{0,n}$ turns out to be
$$
h_{0,n} = \frac{\binom{L-n}{\ell}}{\binom{L}{\ell}}
= \frac{\binom{L-\ell}{n}}{\binom{L}{n}} \simeq \left(1-\frac \ell L\right)^n,
$$
which in practice holds for all types;
in fact, the smallest number of counts, for which the approximation is
better, give the largest contribution to Eq. (\ref{segunda}), 
due to the
power-law
form of $N_L(n)$.
This is given, taking into account a normalisation constant $A$, by
\begin{equation}
N_L(n) = V \frac A {n^\gamma},
\end{equation}
for $n=1,2, \dots$ (and zero otherwise),
with $\sum_{n\ge 1}N_L(n)=V$.  Let us substitute the previous expressions for $h_{0,n}$ and $N_L(n)$ into Eq. (\ref{segunda}), then
\begin{equation}{\label{sum inf}}
v(\ell) \simeq V \left(1-
A\sum_{n\ge 1} \frac{(1-\ell/L)^n}{n^\gamma} \right).
\end{equation}
Although there exists a maximum number of counts $n_{\max}$ beyond which $N_L(n)=0$, 
as a first approximation the sum can be safely extended up to infinity,
and hence we reach the following expression:
\begin{equation}
\label{theory gamma more 2}
v(\ell)\simeq V \left(1-
\frac {\text{Li}_{\gamma}(1-\ell/L)} {\zeta(\gamma)}  \right),\quad
\end{equation}
where we have made use of the polylogarithm function,
$
\text{Li}_{\gamma}(z) = \sum_{n=1}^{\infty} {z^n}/{n^\gamma},$
defined for  $|z| < 1$,
and of the fact that the normalisation of Zipf's law is given by $A=1/\zeta(\gamma)$,
with $\zeta(\gamma)$ the Riemann zeta function,
$\zeta(\gamma)= 
\text{Li}_{\gamma}(1)$.
 Notice that, for random systems with fixed $\gamma$, Eq. (\ref{theory gamma more 2}) yields a
``universal'' scaling relationship between the number of types $v(\ell)$, 
if expressed in units of the total number of types $V$,
and the text position $\ell$ expressed in units of the total size $L$.

In fact, Eq.~\eqref{theory gamma more 2} can lead to an overestimation of $v(\ell)$ due to finite-size effects, but this is rarely noticeable in practice. If one wants a more precise version of Eq.~\eqref{theory gamma more 2}, then, going back to Eq.~\eqref{sum inf} and limiting the sum up to $n_{\max}$ gives, after some algebra,

\begin{equation}
\label{theory gamma less 2}
v(\ell)=V \left( 1 -  \frac
{\text{Li}_{\gamma }(q) - q^{n_{\text{max}}+1} \Phi (q,\gamma ,n_{\text{max}}+1) } 
{\zeta(\gamma) -  \Phi (1,\gamma ,n_{\text{max}}+1) } 
\right),
\end{equation}
with $q=1-\ell/L$, and $\Phi(z,\gamma,a) = \sum_{n=0}^{\infty} \frac{z^n}{(a+n)^\gamma}, \quad |z| < 1; a \neq 0,-1,\dots$ the Lerch transcendent. Obviously, Eq.~\eqref{theory gamma less 2} gives better results at the cost of using an additional parameter, $n_{\max}$. As a rule of thumb, it appears to be worth the cost in cases where $\gamma < 2$, $\ell \ll L$ and $L$ is not too large. 
In most practical cases Eq.~\eqref{theory gamma more 2} gives an excellent approximation;
nevertheless,
we include its more refined version, Eq.~\eqref{theory gamma less 2}, for the sake of completeness.

In order to test these predictions, we simulate a random Zipf's system as follows:
Let us draw $V=10^4$ random numbers $n_1, n_2, \dots n_V$, from the discrete 
probability distribution $N_L(n) /V = n^{-\gamma} / \zeta(\gamma)$, 
with $\gamma=1.8, 2.0$ {and} $2.2$.
Each of these $V$ values of $n$ represents a type, with 
a number of counts given by the value of $n$.
For each type $i=1,\dots,V$, we create then $n_i$ copies (tokens) of its associated type, 
and make a list with all of them,
$$
\underbrace{1,\dots,1}_{n_{1}},\underbrace{2,\dots,2}_{n_{2}}, \dots, \underbrace{V,\dots,V}_{n_{V}}.
$$
Then, the list
is reshuffled in order to create a random system, 
of size $L=n_1+n_2+\dots +n_V$.
Figure \ref{fig random} shows the resulting type-token growth together 
with the approximation given 
either by Eq. \eqref{theory gamma more 2},
which only depends on $\gamma$,
or by Eq. \eqref{theory gamma less 2}, which depends on $\gamma $ and $n_{\text{max}}$.
The agreement is nearly perfect,
except for very small $\ell$.

So far we have shown that Eqs.~\eqref{theory gamma more 2} and \eqref{theory gamma less 2} capture very accurately the type-token growth curve for synthetic systems that have a perfect power-law distribution of counts but are completely random. 
Real systems, however, can have richer structures beyond the distribution of counts
\cite{Manning,Corral_words,Motter}
and so one wonders if our derivations can provide acceptable predictions for
them.

In the following, we show that this is indeed the case when the system considered is that of natural language, and provide a qualitative explanation of this {remarkable} fact.

\begin{figure}
\begin{center}
\includegraphics[width=1.0\columnwidth]{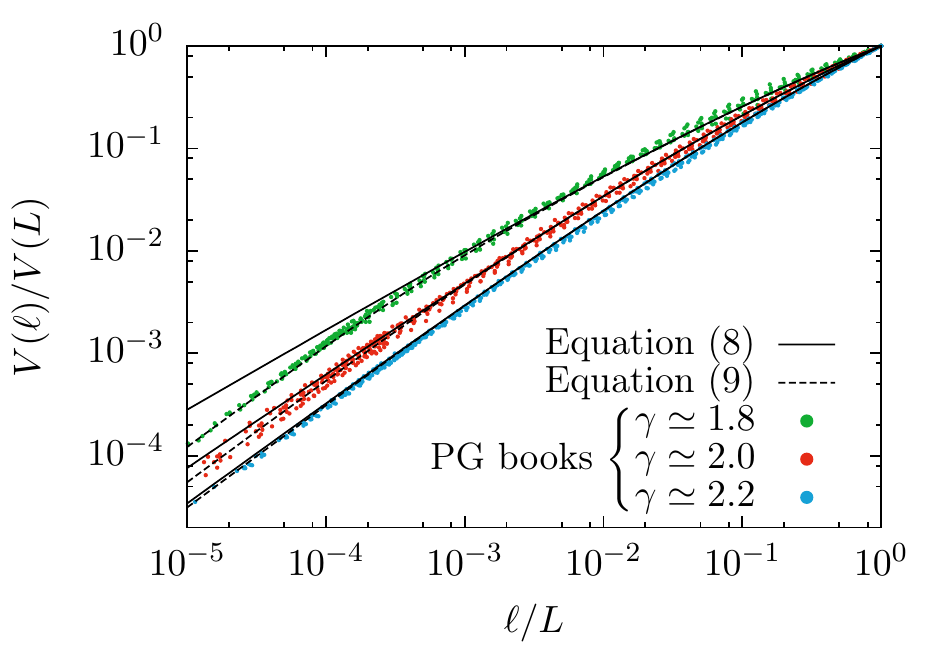}
\end{center}
\caption{
The rescaled vocabulary-growth curve of 28 books from the PG database with exponents $\gamma =\{1.8,2.0,2.2\}\pm 0.01$ 
fitted for $n\ge 1$ or $n\ge 2$.
The values of $L$ and $V$ range from 
$27,873$ to $146,845$
and  from $5,639$ to $30,912$
respectively.
As it is apparent, all books with the same exponent collapse into a single curve, which Eqs.~\eqref{theory gamma more 2} and \eqref{theory gamma less 2} accurately capture. For the case of Eq.~\eqref{theory gamma less 2}, we have used a value of $n_{\max}/L=0.05$.
} 
\label{fig_universal}
\end{figure}

We analyse books from the  \emph{Project Gutenberg} (PG) database \cite{PG},
selecting those whose distribution of frequencies $N_L(n)$ is statistically compatible with a pure, discrete power law distribution.
We fit the $\gamma$ exponent with rigorous methods, see Refs. \cite{Corral_Deluca,Corral_Deluca_arxiv}). 
In analogy with the previous section, 
we plot in Fig.~\ref{fig_universal} $v(\ell)/V$ versus $\ell/L$ 
for a total of 28 books for which $\gamma=1.8,2.0$, or $2.2$. 
Books with the same Zipf's exponent collapse 
between them and into the corresponding theoretical curves,
Eqs.~\eqref{theory gamma more 2} and \eqref{theory gamma less 2}.
This is rather {noticeable}, as it points to the idea that the vocabulary-growth curve is unaffected by clustering, correlations, or by 
syntactic or discursive constraints. 
In other words, the vocabulary-growth curve of a real book {fulfilling Zipf's law as given by Eq. (\ref{Zipfnonorm})} 
is not a power law but
can be predicted using only its associated Zipf's exponent.

In order to understand why a prediction that heavily depends on the randomness hypothesis works so well for real books, we analyse the inter-occurrence-distance
distribution of words. 
Given a word (type) with frequency $n$, we define its $k$-th inter-occurrence distance $\tau_k$ as the number of words (tokens) between its $k-1$-th and $k$-th appearances, plus one;
i.e., 
$$\tau_k=\ell_k-\ell_{k-1}$$ 
(with $\ell_k$ the position of its $k$-th appearance and $k\le n$). 
For the case of $k=1$, we compute the number of words from the beginning of the text
up to the first appearance, i.e., $\tau_1=\ell_1$.
If real books were completely random, then $\tau_k$ would be roughly exponentially distributed,
and the rescaled distances 
\begin{equation}
\label{rescaled waiting}
\hat{\tau}_k = \frac{\tau_k}{\langle \tau_k \rangle}
\end{equation}
would be, for any value of $n$, exponentially distributed with parameter 1. 
Deviations from an exponential distribution for inter-occurrence distances in real books are
 well-known when all $k$ are considered together, 
and constitute the 
so-called clustering or burstiness effect: 
instances of a given word tend to appear grouped together in the book, forming clusters
and hence both very short and very long inter-occurrence distances are much more
common than what an exponential distribution predicts \cite{Corral_words,Motter}. 

\begin{figure}[ht]
\includegraphics[width=.90\columnwidth]{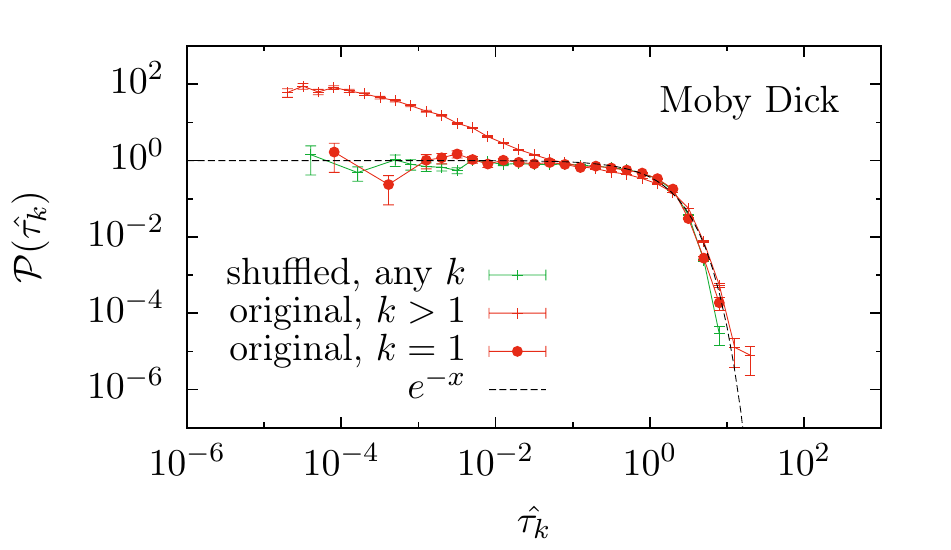}
\includegraphics[width=.90\columnwidth]{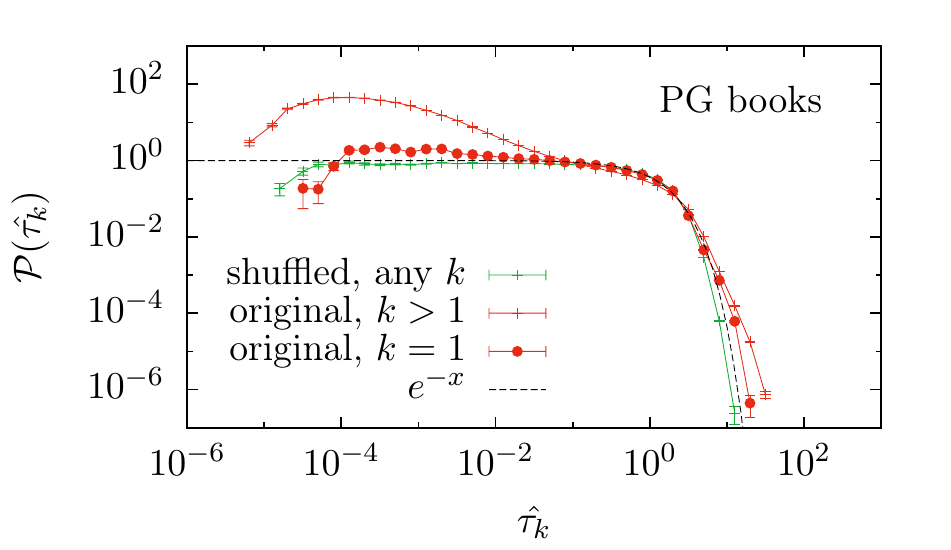}
\caption{Distribution of the rescaled inter-occurrence distances $\hat{\tau}_k$, see Eq.~\eqref{rescaled waiting}. The scale parameter $\langle \tau_k \rangle$ was computed from the data of each book (types with $n=1$ or with $N(n)=1$ were not included in the analysis). The original books (red) display clear deviations from an exponential distribution for $k>1$, but not for $k=1$. Shuffled versions of the books (green) do not show, as expected, any clustering effect, and hence their rescaled inter-occurrence distances are roughly exponentially distributed.
\textbf{Top}: The book \emph{Moby Dick}, by Herman Melville, as an illustrative example. \textbf{Bottom}: The one hundred longest books in the PG database. 
}
\label{fig waiting}
\end{figure}

Our analysis 
introduces an additional
 element, the parameter $k$. Note that for what concerns the vocabulary-growth curve, 
all that matters is $k=1$, as it is only the first appearance of
 each word that adds to the vocabulary. Figure~\ref{fig waiting} shows 
the (estimated) probability mass function $\mathcal{P}(\hat \tau_k)$
of the rescaled inter-occurrence distance 
for the book \emph{Moby Dick} as an example (top), and for the one hundred longest books in the PG database (bottom). As it is
 apparent, for $k > 1$, the distributions of distances are not exponentially
 distributed, and we recover a trace of the clustering effect;
however
 the case $k=1$ displays a clearly different shape, much more close to an
 exponential distribution. This explains, at a qualitative level, why our
 derivations, based on a randomness assumption, continue to work in the case
 of real books that display clustering effects.

In conclusion, we have shown that Eqs.~\eqref{theory gamma more 2}
and \eqref{theory gamma less 2},
which are not power laws but contain the
polylogarithm function and the Lerch transcendent,
provide 
a continuum of universality classes for type-token growth, 
depending only on
Zipf's exponent. We have verified our results both on synthetic random systems and on real
books, showing that despite correlations or clustering effects, they remain valid as long
as Zipf's law is fulfilled. Our results open the door to investigations in other contexts beyond linguistics, where the validity of Heaps' law could be questioned in a similar manner. 

{\it Acknowledgements.} We have benefited from a long-term collaboration
with G. Boleda and R. Ferrer-i-Cancho.
Research projects in which this work is included are
FIS2012-31324, from Spanish MINECO, 
and  2014SGR-1307, from AGAUR.

\addcontentsline{toc}{chapter}{Bibliography}
\bibliography{biblio}

\begin{thebibliography}{10}

\bibitem{Newman_05}
M.~E.~J. Newman.
\newblock Power laws, {Pareto} distributions and {Zipf}'s law.
\newblock {\em Cont. Phys.}, 46:323 --351, 2005.

\bibitem{Clauset}
A.~Clauset, C.~R. Shalizi, and M.~E.~J. Newman.
\newblock Power-law distributions in empirical data.
\newblock {\em SIAM Rev.}, 51:661--703, 2009.

\bibitem{Adamic_Huberman}
L.~A. Adamic and B.~A. Huberman.
\newblock {Zipf's} law and the {Internet}.
\newblock {\em Glottometrics}, 3:143--150, 2002.

\bibitem{Furusawa2003}
C.~Furusawa and K.~Kaneko.
\newblock Zipf's law in gene expression.
\newblock {\em Phys. Rev. Lett.}, 90:088102, 2003.

\bibitem{Axtell}
R.~L. Axtell.
\newblock Zipf distribution of {U.S.} firm sizes.
\newblock {\em Science}, 293:1818--1820, 2001.
\newblock 

\bibitem{Serra_scirep}
J.~Serr\`a, A.~Corral, M.~Bogu{\~n}\'a, M.~Haro, and J.~Ll. Arcos.
\newblock Measuring the evolution of contemporary western popular music.
\newblock {\em Sci. Rep.}, 2:521, 2012.

\bibitem{Malevergne_Sornette_umpu}
Y.~Malevergne, V.~Pisarenko, and D.~Sornette.
\newblock Testing the {Pareto} against the lognormal distributions with the
  uniformly most powerful unbiased test applied to the distribution of cities.
\newblock {\em Phys. Rev. E}, 83:036111, 2011.

\bibitem{Zanette_book}
D.~Zanette.
\newblock {\em Statistical Patterns in Written Language}.
\newblock 2012.

\bibitem{Lu_2010}
L.~L\"u, Z.-K. Zhang, and T.~Zhou.
\newblock Zipf's law leads to {Heaps}' law: Analyzing their relation in
  finite-size systems.
\newblock {\em PLoS ONE}, 5(12):e14139, 12 2010.

\bibitem{Corral_Boleda}
A.~Corral, G.~Boleda, and R.~{Ferrer-i-Cancho}.
\newblock {\em in preparation}, 2013.

\bibitem{Heaps_1978}
H.~S. Heaps.
\newblock {\em Information retrieval: computational and theoretical aspects}.
\newblock Academic Press, 1978.

\bibitem{Baeza_Yates00}
R.~Baeza-Yates and G.~Navarro.
\newblock Block addressing indices for approximate text retrieval.
\newblock {\em J. Am. Soc. Inform. Sci.}, 51(1):69--82, 2000.

\bibitem{Baayen}
H.~Baayen.
\newblock {\em Word Frequency Distributions}.
\newblock Kluwer, Dordrecht, 2001.

\bibitem{Mandelbrot61}
B.~Mandelbrot.
\newblock {On the theory of word frequencies and on related {Markovian} models
  of discourse}.
\newblock In R.~Jakobson, editor, {\em Structure of Language and its
  Mathematical Aspects}, pages 190--219. American Mathematical Society,
  Providence, RI, 1961.

\bibitem{Leijenhorst_2005}
D.C. van Leijenhorst and Th.P. van~der Weide.
\newblock A formal derivation of {Heaps}' law.
\newblock {\em Inform. Sciences}, 170:263 -- 272, 2005.

\bibitem{Kornai2002}
A.~Kornai.
\newblock How many words are there?
\newblock {\em Glottom.}, 2:61--86, 2002.

\bibitem{Serrano}
M.~A. Serrano, A.~Flammini, and F.~Menczer.
\newblock Modeling statistical properties of written text.
\newblock {\em PLoS ONE}, 4(4):e5372, 2009.

\bibitem{FontClos_Corral}
F.~Font-Clos, G.~Boleda, and A.~Corral.
\newblock A scaling law beyond {Zipf}'s law and its relation with {Heaps}' law.
\newblock {\em New J. Phys.}, 15:093033, 2013.
\newblock 

\bibitem{Wimmer_Altmann}
G.~Wimmer and G.~Altmann.
\newblock On vocabulary richness.
\newblock {\em J. Quant. Linguist.}, 6:1--9, 1999.

\bibitem{Sano2012}
Y.~Sano, H.~Takayasu, and M.~Takayasu.
\newblock Zipf's law and {Heaps}' law can predict the size of potential words.
\newblock {\em Prog. Theor. Phys. Supp.}, 194:202--209, 2012.

\bibitem{Bernhardsson_2011}
S.~Bernhardsson, S.~K. Baek, and P.~Minnhagen.
\newblock A paradoxical property of the monkey book.
\newblock {\em J. Stat. Mech.}, 2011(07):P07013, 2011.

\bibitem{Lu_2013}
L.~L{\"u}, Z.-K. Zhang, and T.~Zhou.
\newblock Deviation of {Zipf}'s and {Heaps' Laws} in human languages with
  limited dictionary sizes.
\newblock {\em Sci. Rep.}, 3:1--7, 2013.

\bibitem{Petersen_scirep}
A.~M. Petersen, J.~N. Tenenbaum, S.~Havlin, H.~E. Stanley, and M.~Perc.
\newblock Languages cool as they expand: Allometric scaling and the decreasing
  need for new words.
\newblock {\em Sci. Rep.}, 2:943, 2012.

\bibitem{Gerlach_Altmann}
M.~Gerlach and E.~G. Altmann.
\newblock {Stochastic model for the vocabulary growth in natural languages}.
\newblock {\em Phys. Rev. X}, 3:021006, 2013.

\bibitem{Gerlach}
M.~Gerlach and E.~G. Altmann.
\newblock Scaling laws and fluctuations in the statistics of word frequencies.
\newblock {\em New J. Phys.}, 16(11):113010, 2014.

\bibitem{Corral_words}
A.~Corral, R.~{Ferrer-i-Cancho}, and A.~D\'{\i}az-Guilera.
\newblock Universal complex structures in written language.
\newblock {\em http://arxiv.org}, 0901.2924, 2009.

\bibitem{Motter}
E.~G. Altmann, J.~B. Pierrehumbert, and A.~E. Motter.
\newblock Beyond word frequency: Bursts, lulls, and scaling in the temporal
  distributions of words.
\newblock {\em ArXiv}, 0901.2349v1, 2009.

\bibitem{Manning}
C.~D. Manning and H.~Sch{\"u}tze.
\newblock {\em Foundations of Statistical Natural Language Processing}.
\newblock MIT Press, Cambridge, Massachusetts, 1999.

\bibitem{PG}
{http://www.gutenberg.org/}.

\bibitem{Corral_Deluca}
A.~Deluca and A.~Corral.
\newblock Fitting and goodness-of-fit test of non-truncated and truncated
  power-law distributions.
\newblock {\em Acta Geophys.}, 61:1351--1394, 2013.

\bibitem{Corral_Deluca_arxiv}
A.~Corral, A.~Deluca, and R.~{Ferrer-i-Cancho}.
\newblock A practical recipe to fit discrete power-law distributions.
\newblock {\em ArXiv}, 1209:1270, 2012.

\end{thebibliography}
\bibliographystyle{unsrt}

\end{document}